\documentclass{article}

\usepackage{microtype}
\usepackage{graphicx}
\usepackage{subcaption}
\usepackage{booktabs} 
\usepackage{pgfplots}
\pgfplotsset{compat=1.18}

\usepackage{hyperref}

\usepackage[preprint]{instructions/icml2026}

\usepackage{amsmath}
\usepackage{amssymb}
\usepackage{mathtools}
\usepackage{amsthm}
\usepackage[capitalize,noabbrev]{cleveref}
\usepackage{placeins}

\theoremstyle{plain}

\theoremstyle{definition}

\theoremstyle{remark}

\usepackage[textsize=tiny]{todonotes}

\icmltitlerunning{Protocol Agent [v1]}

\begin{document}

\twocolumn[
  \icmltitle{Protocol Agent: What If Agents Could Use Cryptography In Everyday Life?}

  \begin{icmlauthorlist}
    \icmlauthor{Marco De Rossi}{agent0,metamask}
  \end{icmlauthorlist}

  \icmlaffiliation{agent0}{Agent0}
  \icmlaffiliation{metamask}{MetaMask (Consensys)}

  \icmlcorrespondingauthor{Marco De Rossi}{marco@ag0.xyz}

  \icmlkeywords{cryptography, agentic systems, benchmarks, negotiation, tools}

  \vskip 0.3in
]

\printAffiliationsAndNotice{}  

\begin{abstract}
We often assume that agent-to-agent interaction will mirror human conversation. However, agents operate fundamentally differently. What if they could develop communication patterns that are more efficient and better aligned with their capabilities?

While cryptographic primitives that could profoundly improve everyday interactions already exist, humans can't use them because they are too complex and the math can’t be done in one’s head. Examples range from proving your age (or other attributes) without showing your ID, to filing an anonymous report within a group while proving you are a legitimate member, to splitting a dinner bill fairly without revealing salaries.

What if agents could create protocols ``on the fly'' by recognizing which primitive fits an everyday situation, proposing it to an agentic counterpart, persuading them to participate, and then executing the protocol correctly using appropriate computation tools?

Protocol Agent frames this problem by introducing a benchmark that spans: (1) cryptographic primitive recognition, (2) negotiation skills, (3) implementation correctness, (4) correct computation and (5) security strength. We evaluate current open-weight and state-of-the-art models on this benchmark, propose a dataset-generation approach to improve these capabilities, and measure the impact of supervised fine-tuning (SFT) on benchmark performance, with tuned models outperforming base models by a wide margin.
\end{abstract}

\nocite{derossi2025erc8004}
\nocite{greco2025whatifagents}

\section{Introduction}
\label{sec:introduction}

As agent networks scale, their collective intelligence increasingly depends on a recurring four-step loop:
(i) \emph{comms infrastructure}, how an agent discovers, identifies, and learns to trust other agents;
(ii) \emph{comms standards}, how requests and responses are wrapped and made interoperable;
(iii) \emph{comms behaviors and techniques}, what agents actually say and do, turn by turn, to reach their goals; and
(iv) \emph{learning loops}, how past interactions (their own and others') are turned into improvements via context engineering and post-training.

The first two steps are progressing rapidly. For (i), alongside centralized catalogs and ``stores'' run by large top-of-funnel players, decentralized registries and discovery layers are gaining traction; ERC-8004 \citep{derossi2025erc8004}, for example, has seen growing adoption among startups, researchers, and industry teams building open agent ecosystems. For (ii), standards for tool and agent-to-agent interaction are maturing (e.g., MCP and A2A), while domain-specific protocols are emerging for concrete workflows such as payments (Coinbase's x402) and commerce (Google's AP2 and UCP).

This paper focuses on the third step. We often assume that multi-turn agent interaction will look like human conversation, with the same defaults about disclosure and coordination. Yet agents have very different capabilities from humans, and can execute precise computation on demand, rely on explicit, persistent, machine-readable state (e.g., external memory/tools) and iterate orders of magnitude faster than humans, making feasible coordination behaviors that would not make sense for humans in everyday life.

Our thesis is that agents will not merely automate existing human communication patterns; they will develop interaction protocols that are more efficient and better matched to their capabilities. For example, agents can weave advanced cryptographic primitives into ordinary dialogue (negotiation, compliance flows, purchasing decisions, even ``small talk'') to reduce disclosure, harden agreements, and make outcomes verifiable. Doing this well requires a rare mix of skills: \emph{pattern recognition} (mapping a natural-language situation to a primitive family), \emph{negotiation} (convincing a counterparty to accept a protocol rather than a leaky shortcut), \emph{precise computation/tool use} (producing and reusing correct artifacts), and \emph{security thinking} (anticipating realistic failure modes under social pressure).

A simple example is scheduling: humans typically resolve a meeting by having someone share an availability calendar, implicitly accepting unnecessary leakage. In contrast, \emph{Private Set Intersection} (PSI) has been a mature research topic for over two decades and allows two parties to learn only the overlap of two sets without revealing the rest. This is not something people can do ``in their heads,'' but it is within reach for agents that can compute, negotiate, and produce verifiable artifacts while they coordinate.

The same idea applies to use cases with high economic stakes. For example, agents negotiating the M\&A of two oil companies could identify the regions where both hold advanced reserves without revealing the full extent or the locations of either company's reserves.

Protocol Agent provides (i) an arena to run self-play interactions at scale, (ii) a rubric that decomposes performance into five product-relevant dimensions, and (iii) a training pipeline that supports iterative improvement via tool-grounded conversations, grounded in foundational cryptography references \citep{bonehshoup2023appliedcryptography}.

In our experiments, the tuned models we build with SFT substantially outperform their base counterparts on the benchmark's overall score; for example, \path|deepseek-v3p1| improves from 0.473 to 0.693 (+46.5\%), and \path|qwen3-30b-i2507| improves from 0.390 to 0.676 (+73.3\%).

Finally, Protocol Agent is the first step of a broader open research program on agent comms techniques. To enable community iteration, we release the benchmark, arena, and tool assets as open-source under an MIT license and invite contributions from builders and researchers interested in advancing the project together (see Future work in Section~\ref{sec:limitations}). Source code: \url{https://github.com/agent0lab/protocol-agent}.

\section{Background and Related Work}
\label{sec:background-related}

Cryptographic primitives (e.g., private set intersection, zero-knowledge proofs, signatures, MPC) are typically studied as formal objects with precise security guarantees. Our focus is complementary: we ask whether an agent can deploy these primitives \emph{as social protocols}, under incomplete information, incentives, and objections, and still produce correct, checkable artifacts. This emphasis on application in daily life shifts the bottleneck from derivations to interaction: translating a scenario into a primitive family, negotiating adoption, and executing a procedure with the right checks while minimizing leakage.

Recent benchmarks quantify cryptography capability in LLMs, spanning large-scale QA \citep{elfares2024cryptoqA}, decryption-style reasoning \citep{li2025cipherbank}, and comprehensive suites mixing MCQ/CTF/proof tasks with expert review \citep{wang2025aicrypto}. These are essential for measuring crypto knowledge and reasoning, but they largely abstract away the \emph{coordination layer}, the fact that real deployments require convincing a counterparty and aligning on an interaction protocol.

Parallel lines of work evaluate tool-using agents in security domains where long-horizon planning and execution matter, including CTF benchmarks with LLM-as-judge scoring \citep{shao2025ctfjudge}, runtime-free trajectory synthesis \citep{zhuo2025cyberzero}, broader security/robustness evaluations in real-world tool environments \citep{fu2025raseval}, and secure code generation with objective pipelines (tests + vulnerability checks) \citep{chen2025secureagentbench}. Separately, multi-agent benchmarks study social decision-making and deception dynamics \citep{pan2023machiavelli}, as well as negotiation and deception in hidden-role games \citep{light2023avalonbench}, while ecosystem security frameworks propose cryptographic building blocks for agent identity and policy compliance \citep{adapala2025aegis}. Protocol Agent connects these threads by evaluating whether agents can \emph{select, sell, and execute} privacy-preserving protocols in everyday contexts, where success is jointly social (adoption) and technical (correct artifacts and security reasoning).

\section{Problem Setup}
\label{sec:setup}

\subsection{Threat Model and Assumptions}
\label{sec:threat-model}

\subsubsection{Threat Model.}
Protocol Agent targets everyday interactions where disclosure is costly and incentives are misaligned. Our default threat model is \emph{honest-but-curious} (HBC): a counterparty follows the high-level interaction (they want the stated outcome), but opportunistically infers private information from messages, artifacts, and protocol side effects. Challenges can strengthen or weaken this via role goals/constraints and judge guidance (e.g., active cheating/coercion/replay vs cooperative counterpart); this is encoded primarily via per-role constraints and optional metadata when present.

\subsubsection{Security Objectives.}
We operationalize \emph{Security Strength} by whether the proposed protocol meets the scenario’s confidentiality and integrity goals under the declared threat model. High-scoring solutions (i) minimize leakage beyond what is necessary to achieve the objective, (ii) avoid obvious inference channels (e.g., sending hashed values from a small domain without protections), and (iii) include protocol-level checks that prevent common scenario-relevant attacks (e.g., replay, forgery, selective failure, or coercion-by-oversharing).

\subsubsection{Assumptions and Scope.}
\begin{itemize}
  \item \textbf{Trusted computation tool}: when used, the computation tool (\texttt{cryptomath}) is correct. Attacks on the tool provider, side-channel leakage from tool execution, and compromised endpoints are out of scope.
  \item \textbf{Communication channel}: the transcript is the only channel; we ignore timing/traffic side channels and out-of-band communication.
  \item \textbf{Primitive-family level}: evaluation is at the level of selecting an appropriate \emph{primitive family} and describing a coherent procedure with the right checks. We do not require construction-level proofs; equivalent families are acceptable if they achieve comparable security/leakage for the scenario.
\end{itemize}

\subsection{Task Definition}
\label{sec:task}

Each benchmark instance defines a multi-party interaction with role-specific private information and incentives. The input to a match consists of: (i) a public scenario description, (ii) per-role goals/constraints/private info, and (iii) a symmetry breaker that forces one role to initiate with a required first move (to avoid self-play stalemates).

\subsubsection{Interaction Protocol.}
Participants engage in a turn-based dialogue with fixed role rotation and a maximum number of turns (as configured in the arena runner). Global benchmark rules require that participants (a) keep the setting grounded in everyday life, (b) explicitly name the relevant primitive family at least once, and (c) avoid revealing private inputs unless safe by design and required by the protocol.

\subsubsection{Tooling and Artifacts.}
When a challenge requires computation, participants may call a cryptographic calculator tool to generate at least one \emph{checkable artifact} (e.g., a hash, tag, signature, shared secret, or verifier output) and must integrate the returned value into the protocol step. Tool-usage quality is scored by correctness, necessity, and discipline (no invented outputs; no redundant calls; stop after obtaining the needed artifact).

\subsubsection{Success Criteria and Reported Metrics.}
A successful transcript proposes a concrete, internally consistent protocol that achieves the scenario objective with appropriate checks and minimal leakage, and persuades the counterparty to adopt it. We report five rubric scores (Primitive Selection, Negotiation Skills, Implementation Correctness, Computation / Tool Usage, and Security Strength) and optionally aggregate these into an outcome label for run-level summaries.

\section{Benchmark: Protocol Agent}
\label{sec:benchmark}

Protocol Agent is a benchmark for \emph{deploying} cryptography in everyday interactions. The released benchmark is a single JSON file containing a suite of multi-turn challenges. Each match runs self-play among the specified roles, optionally with access to a computation tool, and is graded by an LLM judge along five dimensions.

Operationally, a benchmark run is: load challenges $\rightarrow$ for each challenge, run $R$ repetitions with fixed budgets (turns, tool calls) $\rightarrow$ collect transcripts and per-dimension scores $\rightarrow$ aggregate into run summaries (means, outcomes, per-challenge breakdowns).

\subsection{Challenge Suite Construction}
\label{sec:challenges}

\textbf{Design goal: primitive-name-free scenarios, primitive-explicit solutions.} The participant-facing scenario and role contexts never name the intended primitive. This forces the agent to map from an everyday description to a primitive family. To make evaluation unambiguous, global rules require that participants explicitly name the primitive family at least once during the discussion, and explain the protocol in plain language.

\textbf{Challenge record format.} Each challenge includes: (i) a set of participants (roles), (ii) a \emph{self-play symmetry breaker} that forces an initiator to make a required first move, (iii) a public background plus per-role goals/constraints/private info, and (iv) judge-only fields: target primitive families, ``what good looks like,'' and common failures. Challenges may also include lightweight validators (e.g., requiring at least one tool result in the transcript for computation-required tasks).

\textbf{Coverage and difficulty.} Challenges are curated to span everyday personal, business, and civic scenarios, with diverse primitive families (e.g., set/attribute protocols, commitments, signatures, authenticated encryption, Merkle proofs). Difficulty labels are coarse (e.g., Intermediate/Hard) and primarily reflect interaction complexity (negotiation + procedure + attack surface), not just cryptographic math.

\begin{center}
\fbox{\begin{minipage}{0.97\linewidth}
\small
\textbf{Everyday $\rightarrow$ primitive (examples).} Representative mappings from the released suite (full catalog in Appendix~\ref{app:challenge-catalog}):
\begin{itemize}
  \item \textbf{``File an anonymous report within a group while proving you are a legitimate member.''} $\rightarrow$ \textbf{Anonymous credentials / group signatures}: prove membership (and optionally properties) without revealing identity.
  \item \textbf{``Find a meeting slot without sharing calendars.''} $\rightarrow$ \textbf{PSI / set-overlap}: reveal only overlap (or a small shortlist), not the full availability pattern.
  \item \textbf{``Prove you are over 21 / income below $X$ without sending ID or tax return.''} $\rightarrow$ \textbf{ZK attribute/range proof}: disclose only the predicate result, not the underlying value.
  \item \textbf{``Check if my password was breached without revealing it, and without enabling scraping.''} $\rightarrow$ \textbf{OPRF-style private membership}: prevent offline guessing/enumeration while enabling a yes/no check.
  \item \textbf{``Buy tokens now, spend later without the issuer linking the two.''} $\rightarrow$ \textbf{Blind signatures}: issuance and redemption become unlinkable while preserving validity and anti-double-spend.
  \item \textbf{``Split a bill fairly without revealing salaries.''} $\rightarrow$ \textbf{MPC}: compute the split from private inputs and reveal only the agreed outputs.
  \item \textbf{``Recover a critical secret only if 2-of-3 trusted people cooperate.''} $\rightarrow$ \textbf{Secret sharing}: distribute trust and eliminate single points of failure.
\end{itemize}
\end{minipage}}
\end{center}

\subsection{Arena: Interaction and Tooling}
\label{sec:arena}

The arena is a deterministic runner that converts each challenge into per-role system prompts, executes a bounded turn-taking loop, and records the full transcript for auditability.

\textbf{System prompts and role views.} For each role, the arena assembles a system prompt containing global participant rules, public background, and that role's goals/constraints/private info, plus the symmetry breaker. During execution, each model call sees the full prior transcript, with other roles' turns presented as ``user'' messages prefixed by the speaker id. After the match, the arena also emits per-role ``views'' of the transcript (what each role saw as user vs assistant), to support debugging and dataset construction.

\textbf{Tool calling protocol.} When enabled, participants can call a computation tool by outputting a single JSON object {\scriptsize\path|{"tool":"cryptomath","op":"...","args":{...}}|}. Tool calls do not consume an extra turn: the arena appends a \texttt{TOOL\_RESULT} message containing the tool's JSON response, and the same role continues. A per-turn safety cap limits tool spamming; tool call parsing is intentionally minimal to work across providers.

\textbf{Streaming and reasoning channels.} The arena uses streaming inference where supported. For providers exposing a separate reasoning channel, hidden reasoning is not logged into transcripts; if a response is truncated into reasoning-only, the arena prompts the model to stop thinking and emit a user-visible final answer (with an $\sim$8k token cap).

\textbf{Run artifacts and logging.} Each run produces: per-match transcripts and judge verdicts, per-role views, and aggregate summaries (means and outcome counts). The runner is compatible with parallel execution across matches and uses fixed seeds to support repeated evaluation.

\subsection{Scoring Dimensions and Judge Design}
\label{sec:judge}

Protocol Agent combines an LLM judge with a deterministic evaluation scaffold. Matches are executed with a fixed turn order and explicit budgets (turns, tool calls); tool calls are parsed and executed by the arena, and tool outputs are injected into the transcript as verbatim \texttt{TOOL\_RESULT} messages. Each match records a \texttt{match\_seed} to support repeated evaluation under the same runner configuration.

\textbf{Structured judge I/O and tool-verified checks.} The judge is prompted with: global judge rules, the global rubric (dimension definitions), the observed tool-result count, and the challenge-specific judge-only fields (target primitives, ``what good looks like,'' common failures). The judge must return a strict JSON object with the five dimension scores (integers 1--5) and a short verdict. When exact computation matters, our reference judge configuration can call \texttt{cryptomath} (with a small cap) and is instructed to recompute/verify any claimed values; this makes many correctness checks deterministic even though overall scoring includes judgment.

\textbf{Dimension semantics (operational).}
\begin{itemize}
  \item \textbf{Primitive Selection}: did participants explicitly name an appropriate primitive family and use it coherently? If no primitive family is ever named, the judge is instructed to assign a 1.
  \item \textbf{Negotiation Skills}: did participants explain benefits/tradeoffs and persuade the counterparty (addressing objections and incentives), rather than demanding trust?
  \item \textbf{Implementation Correctness}: is there a concrete multi-step procedure with the right checks (who computes what, what is revealed, what is verified), consistent end-to-end?
  \item \textbf{Computation / Tool Usage}: when computation is required, are tool calls correct, minimal, and actually used in the protocol? The judge is instructed to check consistency by recomputing with \texttt{cryptomath} when available, penalizing missing, incorrect, or unused artifacts.
  \item \textbf{Security Strength}: does the protocol meet confidentiality/integrity goals under the declared threat model, and avoid common failure patterns listed by the challenge?
\end{itemize}

\subsection{Calculator Tool}
\label{sec:calculator}

\texttt{cryptomath} is an ad hoc, Rust-backed cryptographic calculator we built for this work. Its role is not to ``help the model do crypto,'' but to make protocol steps \emph{checkable}: agents must produce concrete artifacts (hashes, tags, keys, signatures, Merkle proofs, modular arithmetic results), and the arena/judge can deterministically recompute and verify claimed values rather than relying on narrative correctness.

\textbf{Minimal interface.} The tool is exposed via an HTTP endpoint (CLI-compatible JSON protocol, with an AWS Lambda handler in our reference deployment). Requests are \texttt{\{"op":"<op>","args":\{...\}\}} and responses are \texttt{\{"ok":bool,"op":"<op>","result":\{...\}?, "error":\{...\}?, "meta":\{version\}\}}. All byte inputs are passed as 0x-hex strings. Two introspection ops (\texttt{schema} and \texttt{help}) return canonical argument names, output shapes, and examples; the arena encourages models to use these when uncertain.

We intentionally keep the operation surface narrow and scenario-driven. A complete operation list (with args and output fields) is provided in Appendix~\ref{app:cryptomath-api}.

\section{Dataset Generation}
\label{sec:dataset}

We train Protocol Agent with a synthetic, tool-grounded conversation dataset that targets the benchmark’s distinctive end-to-end loop: infer a primitive family from a primitive-name-free everyday scenario, persuade a counterpart to adopt it under social pressure, and execute a coherent procedure with artifacts that can be deterministically checked. Unless otherwise noted, synthetic conversations were generated with \emph{Chat-GPT-5.3}. The core design choice is a \emph{slice-based curriculum}: rather than hoping a single data distribution teaches all subskills, we allocate explicit mass to the failure modes we see in practice and preserve that allocation through postprocessing.\footnote{Slice weights are normalized at build time.}

\textbf{Slices (what each one teaches).} Our default mixture emphasizes execution and tool discipline, not ``talk-only'' crypto:
\begin{itemize}
  \item \textbf{Primitive mapping (22\%)}: recognition from everyday descriptions, with explicit primitive naming and no primitive leakage in the setup.
  \item \textbf{Protocol completion (22\%)}: step-by-step procedures with concrete checks and role responsibilities; computation is tool-grounded when required.
  \item \textbf{Advanced primitives (16\%)}: expands coverage beyond the most common families to improve transfer across the suite.
  \item \textbf{Small-domain pitfalls (10\%)}: supervision on ``tempting shortcuts'' on guessable domains (enumeration/dictionary-style failures) via user-mistake corrections.
  \item \textbf{Tool discipline (12\%)}: trains the distinctive behavior Protocol Agent measures: compute-objective first, minimal calls, reuse-before-regenerate, and stop-after-success.
  \item \textbf{Tool failure drills (10\%)}: one-shot recovery from realistic tool-call errors without collapsing into schema/help loops.
  \item \textbf{Negotiation/persuasion (8\%)}: adoption under objections (``just share the data''), incentive alignment, and plain-language explanation of tradeoffs.
  \item \textbf{Theory anchoring (8\%)}: grounds terminology and security claims in vetted material to reduce invented guarantees.
\end{itemize}

\textbf{Why multiple categories of examples exist.} Within slices we deliberately vary three axes that matter for deployment-quality protocol behavior: (i) \emph{scenario variants} (threat model, interaction constraints, and domain structure) so agents learn conditional choices rather than a single canned recipe; (ii) a \emph{tool decision boundary} (some examples explicitly forbid tool calls, others require them) so agents learn when computation is actually necessary; and (iii) teaching moments where the user proposes an \emph{insecure approach} and the assistant must correct it, turning common failure patterns into explicit supervision.

\textbf{Boneh--Shoup theory distillation.} A unique component of our pipeline is anchoring a fraction of training to a distilled cryptography text. We distilled Boneh and Shoup’s \emph{A Graduate Course in Applied Cryptography} \citep{bonehshoup2023appliedcryptography} into compact chapter-level notes and condition generation on these references. This is not generic ``crypto QA'': the goal is to stabilize \emph{definitions, threat models, and security properties} so that protocol explanations remain consistent when translated into everyday interaction.

\textbf{Deterministic postprocessing to preserve the curriculum.} Tool-grounded transcripts are replayed and validated so artifacts are executable and referenced correctly. Crucially, we avoid naive filtering that would silently delete tool-heavy slices: instead we isolate fragile behaviors (hard ops, error recovery) into dedicated categories and monitor slice/topic survival. Train/validation splits are stratified by slice to keep this curriculum mixture stable across training and evaluation.

\section{Results}
\label{sec:results}

We evaluate models in a multi-turn arena using self-play: for each base model and its tuned counterpart (when available), we run the same suite of challenges under fixed interaction budgets and score the resulting dialogues with a rubric-based judge. This setting stresses end-to-end protocol behavior (selection, negotiation, execution, verification) rather than single-turn crypto knowledge. Full evaluation details (match protocol, budgets, judge configuration, and hardware) are in Appendix~\ref{app:experimental-setup}; training and fine-tuning details are in \cref{sec:dataset} and Appendix~\ref{app:sft-config}.

We report results on the released Protocol Agent suite. All rubric values in this section are normalized to $[0,1]$ via $(s-1)/4$ from the judge’s 1--5 scores; higher is better. Experimental setup details are provided in Appendix~\ref{app:experimental-setup}, and SFT hyperparameters are listed in Appendix~\ref{app:sft-config}.

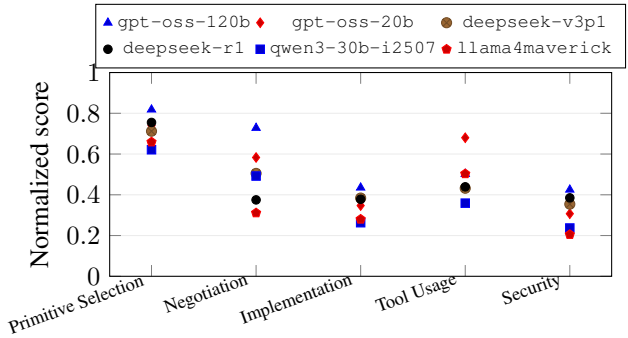
\begin{figure}[t]
\centering
\begin{tikzpicture}
\begin{axis}[
    width=\linewidth,
    height=0.52\linewidth,
    ymin=0, ymax=1,
    ymajorgrids=true,
    grid style={gray!15},
    ylabel={Normalized score},
    symbolic x coords={Primitive,Negotiation,Implementation,Computation,Security},
    xtick=data,
    xticklabels={Primitive Selection,Negotiation,Implementation,Tool Usage,Security},
    xticklabel style={font=\scriptsize,rotate=20,anchor=east},
    legend style={at={(0.5,1.02)},anchor=south,legend columns=3,font=\scriptsize},
    clip=false,
]
\addplot+[only marks, mark=triangle*, mark size=1.8pt] coordinates {
  (Primitive,0.818) (Negotiation,0.728) (Implementation,0.435) (Computation,0.502) (Security,0.425)
};
\addlegendentry{\texttt{gpt-oss-120b}}

\addplot+[only marks, mark=diamond*, mark size=1.8pt] coordinates {
  (Primitive,0.740) (Negotiation,0.583) (Implementation,0.347) (Computation,0.680) (Security,0.307)
};
\addlegendentry{\texttt{gpt-oss-20b}}

\addplot+[only marks, mark=otimes*, mark size=1.9pt] coordinates {
  (Primitive,0.712) (Negotiation,0.505) (Implementation,0.384) (Computation,0.432) (Security,0.354)
};
\addlegendentry{\texttt{deepseek-v3p1}}

\addplot+[only marks, mark=*, mark size=1.6pt] coordinates {
  (Primitive,0.755) (Negotiation,0.375) (Implementation,0.378) (Computation,0.439) (Security,0.385)
};
\addlegendentry{\texttt{deepseek-r1}}

\addplot+[only marks, mark=square*, mark size=1.6pt] coordinates {
  (Primitive,0.621) (Negotiation,0.492) (Implementation,0.263) (Computation,0.359) (Security,0.237)
};
\addlegendentry{\texttt{qwen3-30b-i2507}}

\addplot+[only marks, mark=pentagon*, mark size=1.8pt] coordinates {
  (Primitive,0.659) (Negotiation,0.311) (Implementation,0.280) (Computation,0.503) (Security,0.205)
};
\addlegendentry{\texttt{llama4maverick}}
\end{axis}
\end{tikzpicture}
\caption{Base model rubric profiles on Protocol Agent (normalized). Each model’s five rubric means are plotted as points across dimensions.}
\label{fig:rubric-profile}
\end{figure}

\paragraph{Base model performance.}
Figure~\ref{fig:rubric-profile} shows how base models trade off rubric dimensions (e.g., strong computation but weaker negotiation/security for some models), helping interpret where improvements are needed.

\FloatBarrier

\begin{table}[t]
\centering
\scriptsize
\setlength{\tabcolsep}{2.5pt}
\resizebox{\linewidth}{!}{%
\begin{tabular}{l l r r r r r r}
\toprule
model & type & \rotatebox{60}{Primitive Selection} & \rotatebox{60}{Negotiation} & \rotatebox{60}{Implementation} & \rotatebox{60}{Tool Usage} & \rotatebox{60}{Security} & \rotatebox{60}{Overall} \\
\midrule
\path|deepseek-v3p1-tuned| & tuned & 0.961 & 0.845 & 0.688 & 0.322 & 0.755 & 0.693 \\
\path|qwen3-30b-i2507-tuned| & tuned & 0.917 & 0.785 & 0.629 & 0.444 & 0.639 & 0.676 \\
\path|deepseek-r1-tuned| & tuned & 0.892 & 0.767 & 0.605 & 0.398 & 0.650 & 0.662 \\
\path|gpt-oss-120b-tuned1| & tuned & 0.855 & 0.778 & 0.495 & 0.452 & 0.495 & 0.603 \\
\path|gpt-oss-120b| & base & 0.818 & 0.728 & 0.435 & 0.502 & 0.425 & 0.582 \\
\path|gpt-oss-20b| & base & 0.740 & 0.583 & 0.347 & 0.680 & 0.307 & 0.531 \\
\path|deepseek-v3p1| & base & 0.712 & 0.505 & 0.384 & 0.432 & 0.354 & 0.472 \\
\path|deepseek-r1| & base & 0.755 & 0.375 & 0.378 & 0.439 & 0.385 & 0.457 \\
\path|qwen3-30b-i2507| & base & 0.621 & 0.492 & 0.263 & 0.359 & 0.237 & 0.390 \\
\path|llama4maverick| & base & 0.659 & 0.311 & 0.280 & 0.503 & 0.205 & 0.388 \\
\path|mistral-small-24b-i2501| & base & 0.615 & 0.420 & 0.200 & 0.412 & 0.195 & 0.368 \\
\bottomrule
\end{tabular}
}
\caption{All models on Protocol Agent (normalized rubric means, $[0,1]$). Overall is the mean of the five dimensions.}
\label{tab:all-models}
\end{table}

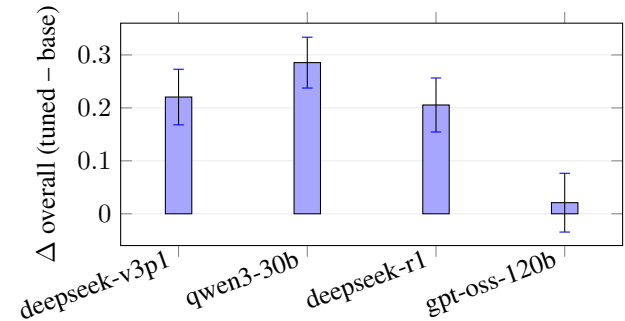
\begin{figure}[b]
\centering
\begin{tikzpicture}
\begin{axis}[
    ybar,
    bar width=10pt,
    width=\linewidth,
    height=0.55\linewidth,
    ymin=-0.06,
    ymax=0.36,
    ymajorgrids=true,
    grid style={gray!15},
    ylabel={$\Delta$ overall (tuned -- base)},
    symbolic x coords={deepseek-v3p1,qwen3-30b,deepseek-r1,gpt-oss-120b},
    xtick=data,
    xticklabel style={rotate=20,anchor=east},
    enlarge x limits=0.15,
]
\addplot+[
    draw=black,
    fill=blue!35,
    error bars/.cd,
    y dir=both,
    y explicit,
] coordinates {
    (deepseek-v3p1,0.2205) +- (0,0.0525)
    (qwen3-30b,0.2855) +- (0,0.0480)
    (deepseek-r1,0.2055) +- (0,0.0510)
    (gpt-oss-120b,0.0210) +- (0,0.0555)
};
\end{axis}
\end{tikzpicture}
\caption{SFT improvement on overall score (normalized). Error bars show the 95\% bootstrap CI half-width from tuned-vs-base comparisons.}
\label{fig:sft-gains}
\end{figure}

\paragraph{SFT drives large gains for some families.}
Across model families where both base and tuned runs exist, SFT yields substantial improvements in end-to-end protocol behavior (see Table~\ref{tab:all-models} and Figure~\ref{fig:sft-gains}). For example, \path|deepseek-v3p1| improves from 0.473 to 0.693 overall ($\Delta=+0.220$), and \path|qwen3-30b-i2507| improves from 0.390 to 0.676 ($\Delta=+0.286$). In contrast, \path|gpt-oss-120b| changes only slightly (0.582 to 0.603; $\Delta=+0.021$).

\paragraph{Gains and remaining gaps.}
The strongest gains concentrate in negotiation and security: tuned models more reliably (i) identify a fitting primitive family from a primitive-name-free scenario, (ii) persuade counterparts away from leaky shortcuts, and (iii) articulate checks against realistic failure modes. Interestingly, some tuned models show \emph{lower} Computation / Tool Usage despite large overall gains (e.g., \path|deepseek-v3p1|: $\Delta=-0.110$, $p\approx 0.020$), suggesting a failure mode where improved ``talk'' outpaces disciplined artifact generation and tool-verification.

\begin{table}[t]
\centering
\small
\setlength{\tabcolsep}{4pt}
\begin{tabular}{lrrrr}
\toprule
Dimension & Base & Tuned & $\Delta$ & $p_{\mathrm{perm}}$ \\
\midrule
Overall (mean of dims) & 0.473 & 0.693 & +0.220 & $\approx 5\times 10^{-4}$ \\
Primitive Selection & 0.712 & 0.961 & +0.249 & $\approx 5\times 10^{-4}$ \\
Negotiation & 0.505 & 0.845 & +0.340 & $\approx 5\times 10^{-4}$ \\
Implementation & 0.384 & 0.688 & +0.304 & $\approx 5\times 10^{-4}$ \\
Tool Usage & 0.432 & 0.322 & -0.110 & 0.020 \\
Security & 0.354 & 0.755 & +0.402 & $\approx 5\times 10^{-4}$ \\
\bottomrule
\end{tabular}
\caption{Deep-dive: \path|deepseek-v3p1| tuned vs base. Scores are normalized ($[0,1]$). $p_{\mathrm{perm}}$ is the deterministic two-sided permutation test p-value over matches (100 per run).}
\label{tab:deepdive-v3p1}
\end{table}

\FloatBarrier
\section{Discussion, Limitations, and Future Work}
\label{sec:limitations}

\subsection{Discussion}
Research on how agents communicate in multi-turn conversations is nascent. Since current models are trained on human-like conversation corpora (or synthetic data that emulates it), their interaction patterns largely reflect human defaults. This behavior will increasingly change as models in the next few years are trained and post-trained with native agentic behavior, and will diverge from human patterns; the use of cryptography in everyday life is just one example of that shift.

The results support our central claim: deploying cryptography in everyday life is not a single skill, but an end-to-end interaction capability. Models that can already reason about primitives often fail in the coordination layer (getting buy-in) or in execution (producing verifiable artifacts and checks). While SFT can bring value and is promising, this is a complex multi-skill capability that will likely require new approaches, including ``unlearning'' corresponding human behaviors that may be inefficient when pursuing agentic goals.

\subsection{Limitations}
Our current evidence has three important limitations:
\begin{itemize}
  \item \textbf{Self-play counterpart bias.} Tuned models interactions use self-play among models that already have substantial cryptography knowledge, which can make primitive recognition less of a bottleneck. Evaluating with counterparts not trained to recognize cryptographic primitives would better isolate negotiation and protocol-selling skills.
  \item \textbf{Model coverage.} We do not yet report results on top state-of-the-art commercial systems (e.g., Claude Opus, ChatGPT, Gemini).
  \item \textbf{Scenario distribution.} The benchmark emphasizes scenarios where cryptographic primitives are a clean fit and are expected to emerge. Performance may differ in more ambiguous, long-running, or less structured conversations where the ``crypto move'' is weaker, delayed, or only partially applicable.
\end{itemize}

\subsection{Future Work}
We see several concrete directions:
\begin{itemize}
  \item \textbf{Reinforcement learning for protocol behavior.} Apply RL techniques (e.g., outcome-based RL, preference optimization, or tool-verified reward shaping) to directly optimize adoption, artifact correctness, and security-strength objectives beyond what static demonstrations capture.
  \item \textbf{Permissionless identity and diversity.} Anchor the arena's participant identity and discovery layer to permissionless agent identity and trust infrastructure (e.g., ERC-8004) to increase agent diversity in an open, censorship-resistant setting.
  \item \textbf{More deterministic validation.} Expand validators and tool-verified checks so that scoring depends less on rubric interpretation and more on replayable verification of protocol steps (not only that a tool was called, but that its outputs are used correctly and verify).
  \item \textbf{Beyond self-play.} Incorporate human-in-the-loop interactions, heterogeneous counterpart pools, and adversarial counterparts to better reflect real deployment conditions.
  \item \textbf{Stronger crypto tooling.} Improve agent capability to use a more advanced cryptography toolchain (richer primitives, better introspection, and safer interfaces), and measure whether stronger tools translate into more reliable, less leaky protocols.
\end{itemize}
\subsection{Takeaways}
Protocol Agent measures whether agents can \emph{select, sell, and execute} cryptography as a coordination product in everyday life. Across multiple families, SFT can produce large, statistically robust gains in end-to-end performance, especially on negotiation and security reasoning. Remaining failures cluster around disciplined artifact generation and verification, a natural target for stronger tool-grounded training and more deterministic validation.

\section*{Acknowledgements}

This work is produced by Agent0 and developed in collaboration with MetaMask.

The idea behind it is inspired by Nicola Greco's talk, ``What If Agents Could Create Protocols on the Fly?'', at ERC-8004's Trustless Agent Day in Buenos Aires.

\bibliography{references}
\bibliographystyle{instructions/icml2026}

\newpage
\appendix
\onecolumn
\section{Fine-Tuning Configuration}
\label{app:sft-config}
This appendix summarizes the supervised fine-tuning (SFT) configuration used for the tuned runs reported in the main text.

\paragraph{Dataset size.}
Across tuned runs, the training and evaluation sets are shared: train pairs = 9294, train tokens = 15{,}319{,}623; eval pairs = 190, eval tokens = 311{,}164.

\medskip
\noindent\begin{center}
\small
\setlength{\tabcolsep}{4pt}
\captionof{table}{SFT hyperparameters for tuned runs.}
\label{tab:sft-hparams}
\begin{tabular}{lrrrrrrr}
\toprule
model & epochs & lr & LoRA rank & warmup & batch & grad accum & weight decay \\
\midrule
\path|deepseek-r1-tuned| & 3 & 1.50e-04 & 32 & 1000 & 32768 & 8 & 0.02 \\
\path|deepseek-v3p1-tuned| & 2 & 1.00e-04 & 32 & 300 & 32768 & 4 & 0.01 \\
\path|gpt-oss-120b-tuned1| & 3 & 1.50e-04 & 32 & 500 & 32768 & 4 & 0.01 \\
\path|qwen3-30b-i2507-tuned| & 2 & 1.00e-04 & 32 & 300 & 32768 & 4 & 0 \\
\bottomrule
\end{tabular}
\end{center}

\section{Experimental Setup}
\label{app:experimental-setup}
We evaluate models in self-play on the released Protocol Agent suite.

\paragraph{Match protocol and budgets.}
Each run consists of 100 self-play matches: 50 challenges, each repeated twice. Matches use turn-based role rotation with a maximum of 15 turns per match. Where supported, the arena uses streaming inference; for providers exposing a separate reasoning channel, hidden reasoning is not logged into transcripts, and if a response is truncated into reasoning-only, the arena prompts the model to stop thinking and emit a user-visible final answer (with an $\sim$8k token cap).

\paragraph{Judge.}
All runs use an LLM judge configured to output strict JSON rubric scores and, when required, to recompute/verify claimed computations using \texttt{cryptomath}, a Rust calculator developed for this project and open-sourced alongside the other artifacts.

\medskip
\noindent\begin{center}
\small
\setlength{\tabcolsep}{4pt}
\captionof{table}{Hardware configurations for the runs reported in this paper.}
\label{tab:hardware}
\begin{tabular}{p{0.58\linewidth}lrr}
\toprule
models & accelerator type & count & replicas \\
\midrule
\path|gpt-oss-120b|, \path|gpt-oss-120b-tuned1|, \path|gpt-oss-20b|, \path|mistral-small-24b-i2501|, \path|qwen3-30b-i2501|, \path|qwen3-30b-i2507|, \path|qwen3-30b-i2507-tuned| & NVIDIA\_H100\_80GB & 2 & 4 \\
\path|deepseek-r1|, \path|deepseek-r1-tuned|, \path|deepseek-v3p1|, \path|deepseek-v3p1-tuned|, \path|llama4maverick| & NVIDIA\_H200\_141GB & 8 & 1 \\
\bottomrule
\end{tabular}
\end{center}

\section{Challenge Catalog by Primitive Family}
\label{app:challenge-catalog}

\noindent This appendix lists all released benchmark challenges grouped by their target primitive family. For each family we give a short application-oriented description and one-line challenge concepts.\par

\subsection*{Anonymous Authentication / Membership}
\noindent\textbf{Primitive concept.} Prove you are an authorized member while keeping your identity hidden (optionally with rate limits).\par
\vspace{2pt}
\begin{itemize}
  \item \textbf{Anonymous Employee Feedback With Proof of Employment} ({\scriptsize\path{anonymous-feedback-with-membership}}): Employee submits anonymous feedback but must prove membership; HR wants optional linkability per epoch for rate limiting.
  \item \textbf{Anonymous Report in a Community Group With Verifiable Membership} ({\scriptsize\path{anonymous-community-report-without-retaliation}}): A community (e.g., apartment building chat, school parents group) wants to allow anonymous reports to moderators, but only from real members. Reporters fear retaliatio...
\end{itemize}

\subsection*{Authenticated Encryption (AEAD)}
\noindent\textbf{Primitive concept.} Encrypt with integrity: confidentiality plus tamper detection, often binding associated metadata to the ciphertext.\par
\vspace{2pt}
\begin{itemize}
  \item \textbf{Secure Software Update: Prevent Rollback and Metadata Swap} ({\scriptsize\path{secure-update-metadata-binding}}): Everyday scenario requiring aead.
  \item \textbf{Share Photos via Cloud: Confidentiality + Tamper Detection + Metadata Binding} ({\scriptsize\path{photo-share-cloud-integrity-metadata}}): Alice wants to share a folder of personal photos with Bob via a cloud provider. Alice wants confidentiality and also wants Bob to detect if the cloud swapped files, re...
\end{itemize}

\subsection*{Authenticated Key Exchange (Out-of-Band)}
\noindent\textbf{Primitive concept.} Establish a shared key with resistance to man-in-the-middle, using a short out-of-band check when available.\par
\vspace{2pt}
\begin{itemize}
  \item \textbf{Detect a Contact Key Change (Prevent Silent MITM)} ({\scriptsize\path{messaging-app-key-change-alert}}): Everyday scenario requiring ake\_oob.
  \item \textbf{Pair Two Devices Securely (Avoid a Man-in-the-Middle)} ({\scriptsize\path{device-pairing-avoid-mitm}}): You're pairing a phone with a new smart device over Bluetooth/Wi‑Fi. A nearby attacker can intercept/modify wireless messages. The device might have only a tiny screen...
\end{itemize}

\subsection*{Blind Signatures}
\noindent\textbf{Primitive concept.} Get a signature on a hidden message so the signer cannot link issuance to later redemption.\par
\vspace{2pt}
\begin{itemize}
  \item \textbf{Anonymous Printing Credits (Issuer Can't Link Purchase to Spend)} ({\scriptsize\path{anonymous-tokens-campus-printing}}): Everyday scenario requiring blind\_signatures.
  \item \textbf{Anonymous Ticket Transfer With Anti-Double-Spend} ({\scriptsize\path{anonymous-event-ticket-transfer-dev}}): Everyday scenario requiring blind\_signatures.
  \item \textbf{Anonymous Tipping Tokens (Platform Can't Link Purchase to Spend)} ({\scriptsize\path{anonymous-tipping-token}}): A platform sells ``tip tokens'' that users can later send to creators. Users want the platform not to be able to link which creator they tipped to their purchase transac...
  \item \textbf{Donate Anonymously But Get a Verifiable Receipt} ({\scriptsize\path{anonymous-donations-blind-sig}}): Everyday scenario requiring blind\_signatures.
  \item \textbf{Transfer an Event Ticket Without Revealing Your Identity to the Venue} ({\scriptsize\path{anonymous-event-ticket-transfer}}): Transfer ticket with anti-double-spend while preserving holder privacy.
\end{itemize}

\subsection*{CCA-Secure Public-Key Encryption}
\noindent\textbf{Primitive concept.} Public-key encryption hardened against active attackers who tamper with ciphertexts.\par
\vspace{2pt}
\begin{itemize}
  \item \textbf{Encrypt Form Fields With Public Headers (Active Attacker)} ({\scriptsize\path{cca-metadata-binding-variant}}): Everyday scenario requiring cca\_pke.
  \item \textbf{Upload an Encrypted Form: Prevent Anyone from Tweaking Fields in Transit} ({\scriptsize\path{anti-malleability-contract-upload}}): An applicant submits an encrypted application form to an online portal. Some metadata must remain public (application type, department), but the sensitive fields (bank...
\end{itemize}

\subsection*{Commitments}
\noindent\textbf{Primitive concept.} Commit now, reveal later: bind to a value while hiding it until you open the commitment.\par
\vspace{2pt}
\begin{itemize}
  \item \textbf{Commit to a Draft Before Peer Review (No After-the-Fact Edits)} ({\scriptsize\path{commit-to-draft-before-review}}): Everyday scenario requiring commitments.
  \item \textbf{Sealed Bid: Commit First, Reveal Later} ({\scriptsize\path{sealed-bid-auction-commitment}}): Everyday scenario requiring commitments.
\end{itemize}

\subsection*{Cryptographic Accumulators}
\noindent\textbf{Primitive concept.} Compact membership proofs over a set, where the proof stays short even if the set is large.\par
\vspace{2pt}
\begin{itemize}
  \item \textbf{Prove You're in a Club Without Revealing Which Member} ({\scriptsize\path{prove-membership-accumulator-variant}}): Everyday scenario requiring accumulator.
  \item \textbf{Prove You're on a Guest List Without Revealing Your Name} ({\scriptsize\path{guest-list-membership-proof}}): An event has a guest list. Door staff needs to verify that a person is on the list, but some guests (e.g., public figures) want privacy and don't want their name spoke...
\end{itemize}

\subsection*{Digital Signatures}
\noindent\textbf{Primitive concept.} Authenticate messages and prevent tampering/repudiation using signatures and challenges.\par
\vspace{2pt}
\begin{itemize}
  \item \textbf{Passwordless Login With Anti-Replay Challenge} ({\scriptsize\path{passwordless-login-avoid-phishing}}): Everyday scenario requiring digital\_signatures.
  \item \textbf{Secondhand Sale: Signed Receipt to Prevent Later Disputes} ({\scriptsize\path{secondhand-sale-signed-receipt}}): You're selling a used item (e.g., laptop, bicycle). Weeks later, disputes can arise (``You never paid,'' ``You never delivered,'' ``That wasn't the agreed price''). You want...
\end{itemize}

\subsection*{Garbled Circuits (2PC)}
\noindent\textbf{Primitive concept.} A practical 2-party computation technique: one party garbles a circuit and the other evaluates privately.\par
\vspace{2pt}
\begin{itemize}
  \item \textbf{Private Health Compatibility Test (Only Learn Yes/No)} ({\scriptsize\path{private-health-compatibility-test}}): Two people want to compute a compatibility/eligibility decision (yes/no) based on private medical/health markers (e.g., genetic marker match, medication interaction)....
\end{itemize}

\subsection*{Group Encryption / Revocation}
\noindent\textbf{Primitive concept.} Group access control and revocation without re-encrypting everything for all remaining members.\par
\vspace{2pt}
\begin{itemize}
  \item \textbf{Shared Album: Revoke One Person Without Re-encrypting Everything} ({\scriptsize\path{shared-photo-album-revocation}}): Everyday scenario requiring group\_encryption.
  \item \textbf{Shared Subscription: Remove an Ex-Roommate's Access Without Re-Encrypting Everything} ({\scriptsize\path{shared-subscription-revocation}}): A household shares access to an encrypted content feed or shared drive. Someone moves out and must lose access immediately. The owner wants revocation without re-encry...
\end{itemize}

\subsection*{Hash-Based Signatures}
\noindent\textbf{Primitive concept.} Conservative signature families relying mainly on hash functions for long-term robustness.\par
\vspace{2pt}
\begin{itemize}
  \item \textbf{Long-Term Document Signing With Conservative Assumptions} ({\scriptsize\path{hash-based-signatures-archive-variant}}): Everyday scenario requiring hash\_based\_signatures.
\end{itemize}

\subsection*{Merkle Transparency / Append-Only Logs}
\noindent\textbf{Primitive concept.} Append-only logs and Merkle proofs to make edits auditable and prevent silent history rewrites.\par
\vspace{2pt}
\begin{itemize}
  \item \textbf{Append-Only Transparency Log for Group Decisions (No Secret Edits)} ({\scriptsize\path{append\_only\_transparency\_for\_group\_decisions}}): A group (club, HOA, open-source project) wants a public record of decisions and votes, but worries the organizer could silently edit or delete inconvenient past entrie...
  \item \textbf{Cloud Folder: Signed Manifest to Detect Deletions/Replays} ({\scriptsize\path{secure-file-share-manifest}}): Everyday scenario requiring merkle\_transparency.
  \item \textbf{Long-Term Family Archive Authenticity (Post-Quantum-Friendly Signatures)} ({\scriptsize\path{post-quantum-family-archive-signing}}): A family wants to store a long-term digital archive (letters, wills, photos) so that decades later, descendants can verify which documents were authentic and unmodifie...
  \item \textbf{Public Contract Version Log (Detect Equivocation)} ({\scriptsize\path{public-log-for-contract-versions}}): Everyday scenario requiring merkle\_transparency.
\end{itemize}

\subsection*{OPRF-Style Private Membership}
\noindent\textbf{Primitive concept.} Private membership checks (e.g., breach or eligibility) designed to resist offline guessing/scraping.\par
\vspace{2pt}
\begin{itemize}
  \item \textbf{Check if Your Password Was Breached Without Revealing It} ({\scriptsize\path{password-breach-check-private}}): A user wants to check whether their password appears in a breached password dataset. They don't want to reveal their password to the service. The service doesn't want...
  \item \textbf{Private Breach Check for Email Address (Service Prevents Enumeration)} ({\scriptsize\path{private-breach-check-oprf-variant}}): Everyday scenario requiring oprf\_private\_membership.
  \item \textbf{Private Coupon Eligibility Check With Anti-Scraping} ({\scriptsize\path{private-coupon-eligibility-oprf}}): Everyday scenario requiring oprf\_private\_membership.
\end{itemize}

\subsection*{Oblivious Transfer (OT)}
\noindent\textbf{Primitive concept.} Let a receiver learn exactly one of two secrets without revealing which they chose.\par
\vspace{2pt}
\begin{itemize}
  \item \textbf{Access One of Two Documents Without Revealing Which} ({\scriptsize\path{pick-one-secret-document-ot}}): Everyday scenario requiring oblivious\_transfer.
  \item \textbf{Choose One of Two Discounts Without Revealing Which} ({\scriptsize\path{choose-one-of-two-discounts-ot}}): Everyday scenario requiring oblivious\_transfer.
  \item \textbf{Choose One of Two Secrets Without Revealing Which One You Chose} ({\scriptsize\path{choose-one-recipe-without-revealing-choice}}): A chef friend offers to share exactly one of two secret recipes (Recipe A or Recipe B). The receiver wants to keep their preference private, and the chef wants to ensu...
\end{itemize}

\subsection*{Onion Routing}
\noindent\textbf{Primitive concept.} Route messages through layers so intermediaries cannot easily link sender and recipient.\par
\vspace{2pt}
\begin{itemize}
  \item \textbf{Send a Sensitive Message Without Revealing Where It Came From} ({\scriptsize\path{anonymous-routing-for-sensitive-message}}): A sender wants to message a recipient without revealing to a network observer who is talking to whom. The network is monitored; some relays may be malicious. The sende...
  \item \textbf{Whistleblower Tip Line: Hide Sender-to-Recipient Link} ({\scriptsize\path{onion-routing-whistleblower-variant}}): Everyday scenario requiring onion\_routing.
\end{itemize}

\subsection*{Optimistic Fair Exchange}
\noindent\textbf{Primitive concept.} Fair exchange with an arbiter used only on disputes: either both sides get what they want or it resolves.\par
\vspace{2pt}
\begin{itemize}
  \item \textbf{Fair Exchange: Digital Art for Payment (Optimistic Arbiter)} ({\scriptsize\path{fair-exchange-digital-art}}): Everyday scenario requiring optimistic\_fair\_exchange.
  \item \textbf{Freelance Work: Fair Exchange of Deliverable for Payment (Optimistic Arbiter)} ({\scriptsize\path{freelance-fair-deliverable-exchange}}): A freelancer is delivering a valuable digital asset (source code, design files). Client will pay, but both fear cheating: client fears paying and receiving nothing/jun...
\end{itemize}

\subsection*{Private Keyword Test}
\noindent\textbf{Primitive concept.} Let a server test for trigger words or patterns without learning full content (or without learning the keywords).\par
\vspace{2pt}
\begin{itemize}
  \item \textbf{Private Keyword Alerts on Messages (Server Can Only Test Triggers)} ({\scriptsize\path{private-trigger-words-variant}}): Everyday scenario requiring keyword\_test.
  \item \textbf{Private Trigger-Word Alerts Without Revealing Full Subject} ({\scriptsize\path{private-email-trigger-words}}): A recipient wants an email server to forward messages to their phone only if the subject contains certain trigger words (e.g., ``urgent'', ``school''). The server must not...
\end{itemize}

\subsection*{Private Set Intersection (PSI)}
\noindent\textbf{Primitive concept.} Compute the overlap between two sets without revealing non-overlapping elements.\par
\vspace{2pt}
\begin{itemize}
  \item \textbf{Deduplicate Contacts With a Friend Without Revealing Your Full Contact List} ({\scriptsize\path{contacts-dedup-private}}): Alice and Bob want to find overlapping contacts (emails/phones) to avoid duplicates, without revealing non-overlapping contacts.
  \item \textbf{Find Common Free Time Without Sharing Calendars} ({\scriptsize\path{personal-calendar-overlap-without-sharing}}): Alice and Bob want to schedule a meeting. Bob suggests: ``Just send me your full calendar and I'll pick a slot.'' Alice wants privacy and only wants to reveal the final...
  \item \textbf{Mutual Likes Without Revealing Who Else You Liked} ({\scriptsize\path{dating-app-match-without-revealing-likes}}): Two users want to learn only whether they mutually liked each other without revealing other likes.
\end{itemize}

\subsection*{Secret Sharing / Threshold Recovery}
\noindent\textbf{Primitive concept.} Split a secret so only a threshold of shares can reconstruct it (e.g., 2-of-3 recovery).\par
\vspace{2pt}
\begin{itemize}
  \item \textbf{Family Emergency Vault: Recover a Secret Only With 2-of-3 Trusted People} ({\scriptsize\path{family-emergency-vault-threshold}}): A person wants to store an emergency secret (e.g., password manager recovery key, important documents). They want a scheme where any 2 of 3 trusted people can help rec...
  \item \textbf{Split a Recovery Key Across Devices (2-of-2 Required)} ({\scriptsize\path{split-key-between-devices}}): Everyday scenario requiring secret\_sharing.
\end{itemize}

\subsection*{Secure Multi-Party Computation (MPC)}
\noindent\textbf{Primitive concept.} Compute a function jointly without revealing each party's private inputs beyond what the output implies.\par
\vspace{2pt}
\begin{itemize}
  \item \textbf{Compute a Fraud Score From Two Companies Without Sharing Raw Data} ({\scriptsize\path{mpc-fraud-check}}): Everyday scenario requiring mpc.
  \item \textbf{Private Group Contribution Computation (Integrity High Level)} ({\scriptsize\path{group-expenses-verify-contributions}}): Everyday scenario requiring mpc.
  \item \textbf{Split Shared Expenses Fairly Without Revealing Incomes} ({\scriptsize\path{split-expenses-without-revealing-incomes}}): Three friends plan a trip and agree to split costs proportionally to income. Nobody wants to reveal their exact income to the others. They want to compute contribution...
\end{itemize}

\subsection*{Zero-Knowledge Attribute/Range Proofs}
\noindent\textbf{Primitive concept.} Prove an attribute predicate (e.g., age over 21, income below X) without revealing the underlying value.\par
\vspace{2pt}
\begin{itemize}
  \item \textbf{Prove You Live in the City Without Revealing Your Exact Address} ({\scriptsize\path{prove-residency-without-address}}): Prove residency eligibility without revealing exact address.
  \item \textbf{Prove Your Income Is Below a Threshold Without Sharing Your Tax Return} ({\scriptsize\path{prove-income-below-threshold-without-tax-return}}): Agency requires income below a threshold. Applicant wants to prove eligibility without revealing full documents.
  \item \textbf{Prove You're Over 21 Without Showing Your ID} ({\scriptsize\path{prove-age-without-showing-id}}): A venue requires proof that you're over 21. The staff asks for your full ID and wants to photocopy it ``for compliance.'' You're willing to prove eligibility but not dis...
\end{itemize}

\section{Cryptomath Tool API}
\label{app:cryptomath-api}
%
%
%

\paragraph{Operations.}
\begin{description}
  \item[\texttt{ping}] args \texttt{\{\}} $\rightarrow$ result \texttt{\{pong:string, version:string\}}.
  \item[\texttt{schema}] args \texttt{\{op:string\}} $\rightarrow$ result \texttt{\{op, expected\_args\_schema, expected\_output\_shape, example\_args\}}.
  \item[\texttt{help}] args \texttt{\{op:string\}} $\rightarrow$ result \texttt{\{op, usage, expected\_args\_schema, expected\_output\_shape, example\_args\}}.

  \item[\texttt{sha256}] args \texttt{\{data:bytes\}} $\rightarrow$ result \texttt{\{digest\_hex:bytes(32)\}}.
  \item[\texttt{sha512}] args \texttt{\{data:bytes\}} $\rightarrow$ result \texttt{\{digest\_hex:bytes(64)\}}.
  \item[\texttt{sha256\_chain}] args \texttt{\{data:bytes, iters:int\}} $\rightarrow$ result \texttt{\{digest\_hex:bytes(32)\}}.
  \item[\texttt{hmac\_sha256}] args \texttt{\{key:bytes, msg:bytes\}} $\rightarrow$ result \texttt{\{tag\_hex:bytes(32)\}}.
  \item[\texttt{hkdf\_sha256}] args \texttt{\{ikm:bytes, salt:bytes, info:bytes, len:int\}} $\rightarrow$ result \texttt{\{okm\_hex:bytes\}}.

  \item[\texttt{aes\_gcm\_encrypt}] args \texttt{\{key:bytes(32), nonce:bytes(12), aad:bytes, plaintext:bytes\}} $\rightarrow$ result \texttt{\{ciphertext\_hex:bytes\}}.

  \item[\texttt{x25519\_keygen}] args \texttt{\{\}} $\rightarrow$ result \texttt{\{sk:bytes(32), pk:bytes(32)\}}.
  \item[\texttt{x25519\_dh}] args \texttt{\{sk:bytes(32), pk:bytes(32)\}} $\rightarrow$ result \texttt{\{shared:bytes(32)\}}.

  \item[\texttt{ed25519\_keygen}] args \texttt{\{\}} $\rightarrow$ result \texttt{\{sk:bytes(32), pk:bytes(32)\}}.
  \item[\texttt{ed25519\_sign}] args \texttt{\{sk:bytes(32), msg:bytes\}} $\rightarrow$ result \texttt{\{sig:bytes(64)\}}.
  \item[\texttt{ed25519\_verify}] args \texttt{\{pk:bytes(32), msg:bytes, sig:bytes(64)\}} $\rightarrow$ result \texttt{\{ok:bool\}}.

  \item[\texttt{secp256k1\_keygen}] args \texttt{\{\}} $\rightarrow$ result \texttt{\{sk:bytes(32), pk:bytes(33)\}}.
  \item[\texttt{secp256k1\_add}] args \texttt{\{p1:bytes(33), p2:bytes(33)\}} $\rightarrow$ result \texttt{\{p\_out:bytes(33)\}}.
  \item[\texttt{secp256k1\_scalar\_mul}] args \texttt{\{p:bytes(33), scalar:bytes(32)\}} $\rightarrow$ result \texttt{\{p\_out:bytes(33)\}}.
  \item[\texttt{secp256k1\_pedersen\_commit}] args \texttt{\{value:bytes(32), blind:bytes(32)\}} $\rightarrow$ result \texttt{\{commitment:bytes(33), H:bytes(33)\}}.
  \item[\texttt{secp256k1\_schnorr\_sign}] args \texttt{\{sk:bytes(32), msg:bytes\}} $\rightarrow$ result \texttt{\{sig:bytes(64), pk\_xonly:bytes(32)\}}.
  \item[\texttt{secp256k1\_schnorr\_verify}] args \texttt{\{pk\_xonly:bytes(32), msg:bytes, sig:bytes(64)\}} $\rightarrow$ result \texttt{\{ok:bool\}}.

  \item[\texttt{zk\_age\_over\_21\_prove}] args \texttt{\{cred\_sk:bytes(32), nonce:bytes\}} $\rightarrow$ result \texttt{\{pk\_xonly:bytes(32), proof\_sig:bytes(64)\}}.
  \item[\texttt{zk\_age\_over\_21\_verify}] args \texttt{\{pk\_xonly:bytes(32), nonce:bytes, proof\_sig:bytes(64)\}} $\rightarrow$ result \texttt{\{ok:bool\}}.

  \item[\texttt{modexp}] args \texttt{\{base:int, exp:int, mod:int\}} $\rightarrow$ result \texttt{\{value\_hex:int\}}.
  \item[\texttt{invmod}] args \texttt{\{a:int, mod:int\}} $\rightarrow$ result \texttt{\{value\_hex:int\}}.
  \item[\texttt{addmod}] args \texttt{\{a:int, b:int, mod:int\}} $\rightarrow$ result \texttt{\{value\_hex:int\}}.
  \item[\texttt{mulmod}] args \texttt{\{a:int, b:int, mod:int\}} $\rightarrow$ result \texttt{\{value\_hex:int\}}.
  \item[\texttt{gcd}] args \texttt{\{a:int, b:int\}} $\rightarrow$ result \texttt{\{value\_hex:int\}}.
  \item[\texttt{crt}] args \texttt{\{residues:array<int>, moduli:array<int>\}} $\rightarrow$ result \texttt{\{x\_hex:int, modulus\_hex:int\}}.

  \item[\texttt{merkle\_parent\_sha256}] args \texttt{\{left:bytes, right:bytes\}} $\rightarrow$ result \texttt{\{digest\_hex:bytes(32)\}}.
  \item[\texttt{merkle\_verify\_path\_sha256}] args \texttt{\{leaf:bytes, siblings:array<bytes>, index:int, root:bytes\}} $\rightarrow$ result \texttt{\{computed\_root\_hex:bytes(32), valid:bool\}}.

  \item[\texttt{bls\_keygen}] args \texttt{\{\}} $\rightarrow$ result \texttt{\{sk\_hex:bytes(32), pk\_hex:bytes(96), variant:string\}}.
  \item[\texttt{g1\_hash\_to\_curve}] args \texttt{\{msg:bytes\}} $\rightarrow$ result \texttt{\{bytes\_hex:bytes(48)\}}.
  \item[\texttt{pairing\_product\_check}] args \texttt{\{terms:array\{g1:bytes(48), g2:bytes(96), sign:1|-1\}\}} $\rightarrow$ result \texttt{\{ok:bool\}}.

  \item[\texttt{rsa\_keygen}] args \texttt{\{bits:int\}} $\rightarrow$ result \texttt{\{n\_hex:int, e\_hex:int, d\_hex:int, p\_hex:int, q\_hex:int\}}.
\end{description}

\end{document}